\documentclass[10pt,a4paper,oneside,onecolumn,preprint,number,sort,compress]{elsarticle}
\pdfoutput=1
\usepackage{amsmath}
\usepackage{amssymb}
\usepackage{subfigure}
\usepackage{hyperref}
\usepackage{mathrsfs}

\newcommand{\BBdmix}{$B^0_d$--$\bar B^0_d$}

\newcommand{\KKmix}{$K^0$--$\bar K^0$}

\newcommand{\KET}[1]{|#1\rangle}
\newcommand{\BRA}[1]{\langle#1|}
\newcommand{\BRAKET}[2]{\langle#1|#2\rangle}
\newcommand{\nof}[1]{P_{\!\!\nrightarrow#1}}
\newcommand{\noft}[1]{\nof{#1}(t)}
\newcommand{\Knof}[1]{\KET{\nof{#1}}}

\newcommand{\Knoft}[1]{\KET{\noft{#1}}}

\newcommand{\nofp}[1]{P_{\!\!\nrightarrow#1}^\perp}

\newcommand{\Knofp}[1]{\KET{\nofp{#1}}}

\newcommand{\Amp}[1]{A_{#1}}
\newcommand{\AmpB}[1]{\bar{A}_{#1}}

\newcommand{\transPROB}[2]{\text{Pr}[{#1}\to {#2}]}

\newcommand{\Li}[1]{\lambda_{#1}}

\newcommand{\Ci}[1]{C_{#1}}
\newcommand{\Si}[1]{S_{#1}}
\newcommand{\Ri}[1]{R_{#1}}
\newcommand{\Cf}{\Ci{f}}
\newcommand{\Sf}{\Si{f}}
\newcommand{\Rf}{\Ri{f}}
\newcommand{\Cg}{\Ci{g}}
\newcommand{\Sg}{\Si{g}}
\newcommand{\Rg}{\Ri{g}}

\newcommand{\tCh}{\cosh\left(\frac{\Delta\Gamma\,t}{2}\right)}
\newcommand{\tCc}{\cos(\Delta m\,t)}
\newcommand{\tSh}{\sinh\left(\frac{\Delta\Gamma\,t}{2}\right)}
\newcommand{\tSc}{\sin(\Delta m\,t)}
\newcommand{\srvC}[2]{\mathcal C_{#1}[#2]}
\newcommand{\srvS}[2]{\mathcal S_{#1}[#2]}
\newcommand{\trnC}[2]{\mathscr C_{#1}[#2]}
\newcommand{\trnS}[2]{\mathscr S_{#1}[#2]}
\newcommand{\srvCh}[1]{\srvC{h}{#1}}
\newcommand{\srvCc}[1]{\srvC{c}{#1}}
\newcommand{\srvSh}[1]{\srvS{h}{#1}}
\newcommand{\srvSc}[1]{\srvS{c}{#1}}
\newcommand{\trnCh}[2]{\trnC{h}{#1,#2}}
\newcommand{\trnCc}[2]{\trnC{c}{#1,#2}}
\newcommand{\trnSh}[2]{\trnS{h}{#1,#2}}
\newcommand{\trnSc}[2]{\trnS{c}{#1,#2}}
\newcommand{\numC}[2]{\mathbf{C}^N[#1,#2]}
\newcommand{\numS}[3]{\mathbf{S}_{#1}^N[#2,#3]}
\newcommand{\numSh}[2]{\numS{h}{#1}{#2}}
\newcommand{\numSc}[2]{\numS{c}{#1}{#2}}

\newcommand{\Pmeson}[1]{#1^0}
\newcommand{\Pbarmeson}[1]{\bar #1^0}
\newcommand{\Pket}[1]{\KET{\Pmeson{#1}}}
\newcommand{\Pbarket}[1]{\KET{\Pbarmeson{#1}}}

\newcommand{\re}[1]{\text{Re}\!\left(#1\right)}
\newcommand{\im}[1]{\text{Im}\!\left(#1\right)}
\newcommand{\Abs}[1]{\left|{#1}\right|}
\newcommand{\abs}[1]{|{#1}|}

\newcommand{\refeq}[1]{(\ref{#1})}
\newcommand{\eq}[1]{eq.\refeq{#1}}

\newcounter{notas}
\begin{document}

\begin{frontmatter}
\title{Novel $T$-Violation observable open to any pair of decay channels at meson factories}
\author[val]{Jos\'e Bernab\'eu}\ead{Jose.Bernabeu@uv.es}
\author[val]{Francisco J. Botella}\ead{Francisco.J.Botella@uv.es}
\author[val]{Miguel Nebot}\ead{Miguel.Nebot@uv.es}
\address[val]{Departament de F\' \i sica Te\`orica and IFIC,Universitat de Val\`encia -- CSIC,\\  E-46100, Burjassot, Spain}

%
\begin{abstract}
Quantum Entanglement between the two neutral mesons produced in meson factories has allowed the first indisputable direct observation of Time Reversal Violation in the time evolution of the neutral meson between the two decays. The exceptional meson transitions are directly connected to semileptonic and CP-eigenstate decay channels. The possibility of extending the observable asymmetries to more decay channels confronts the problem of the ``orthogonality condition'', which can be stated with this tongue-twister: Given a decay channel $f$, Which is the decay channel $f'$ such that the meson state not decaying to $f'$ is orthogonal to the meson state not decaying to $f$? In this paper we propose an alternative $T$-Violation Asymmetry at meson factories which allows its opening to any pair of decay channels. Instead of searching which is the pair of decay channels associated to the $T$-reverse meson transition, we build an asymmetry which tags the initial states of both the Reference and the $T$-reverse meson transitions. This observable filters the appropriate final states by means of two measurable survival probabilities. We discuss the methodology to be followed in the analysis of the new observable and the results expected in specific examples.
\end{abstract}
%
%
\end{frontmatter}
\newpage
The BABAR Collaboration has recently reported, in the \BBdmix\ system, the first direct observation of $T$-violation \cite{Lees:2012uka} in the time evolution of any system, with high statistical significance. The measurement is based in a method described in \cite{Bernabeu:2012ab} following the concepts originally proposed in \cite{Banuls:1999aj,Banuls:2000ki}. The two quantum effects of \emph{entanglement} between the two neutral mesons produced at a meson factory and the \emph{filtering measurement} induced by the meson decay, have to be incorporated in the analysis for the preparation of initial and final meson states.

In the $\left\{\Pket{P},\Pbarket{P}\right\}$ system, where $\Pmeson{P}$ stands for a neutral meson $\Pmeson{K}$, $\Pmeson{D}$, $\Pmeson{B_d}$, or $\Pmeson{B_s}$ (and $\Pbarmeson{P}$ for the corresponding antimeson), we define two arbitrary states:
\begin{eqnarray*}
\KET{P_{1}} &=&p_{1}^{0}\,\Pket{P}+\overline{p}_{1}^{0}\,\Pbarket{P}\,,\\
\KET{P_{2}} &=&p_{2}^{0}\,\Pket{P}+\overline{p}_{2}^{0}\,\Pbarket{P}\,.
\end{eqnarray*}%
The $T$-Violation observable is then given by
\begin{equation}
A_{12}(t)=\frac{\transPROB{P_1}{P_2(t)}-\transPROB{P_2}{P_1(t)}}{\transPROB{P_1}{P_2(t)}+\transPROB{P_2}{P_1(t)}}\,,\label{eq:asymmetry00}
\end{equation}
where 
\begin{equation*}
\transPROB{P_1}{P_2(t)}=\abs{\BRAKET{P_2}{P_1(t)}}^2
\end{equation*}%
is the probability that an initially prepared state $P_{1}$, evolving after time $t$, with the two-state evolution operator $U_{2}(t,0)$, to $P_{1}(t)$, is the state $P_{2}$. In \cite{Lees:2012uka,Bernabeu:2012ab,Banuls:1999aj,Banuls:2000ki} one has four asymmetries \refeq{eq:asymmetry00} built with $P_{1}=\Pmeson{B_{d}},\Pbarmeson{B_{d}}$ and $P_{2}=B_{\pm}$ where $B_{\pm}$ are the $B$ states filtered by the decays to $CP$ eigenstates with definite flavour content. These asymmetries are experimentally independent of $CP$ violation. The Kabir asymmetry \cite{Kabir:1970ts}, eq. (\ref{eq:asymmetry00}) with $P_{1}=\Pmeson{K}$ and $P_{2}=\Pbarmeson{K}$, was measured by the CPLEAR collaboration with a non-vanishing value close to four standard deviations significance \cite{Angelopoulos:1998dv}. But its interpretation in terms of genuine $T$ violation generated controversy \cite{Wolfenstein:1999xb} due to the fact that the observable needs, through $\Gamma _{12}$, the decaying character as an essential ingredient.

Suppose we take, as a Reference, a transition between $P$ states -- some particular combination of $\Pmeson{P}$ and $\Pbarmeson{P}$ -- associated to a given pair of decays: $f$ at $t_{1}$ and $g$ at $t_{2}>t_{1}$. The entangled state of the two mesons in an antisymmetric combination of individual orthogonal states, tells us that the (still living) meson at time $t_{1}$ is tagged, up to a global phase, as ``the state that does not decay into $f$''\cite{Lipkin:1988fu,Branco:1999fs} 
\begin{equation}
\Knof{f} =\frac{1}{\sqrt{\Abs{\Amp{f}}^{2}+\Abs{\AmpB{f}}^{2}}}\Big[\AmpB{f}\Pket{P} -\Amp{f}\Pbarket{P} \Big] \,,
\end{equation}
where $\Amp{f}$ ($\AmpB{f}$) is the decay amplitude from $\Pmeson{P}$ ($\Pbarmeson{P}$) to $f$.  For hadronic decays we are assuming a two body decay channel with a single helicity amplitude, i.e. with spin quantum numbers $0\to j+0$. The corresponding orthogonal state $\BRAKET{\nofp{f}}{\nof{f}} =0$ is -- again up to a global phase -- given by
\begin{equation}
\Knofp{f} =\frac{1}{\sqrt{\Abs{\Amp{f}}^{2}+\Abs{\AmpB{f}}^{2}}}\Big[\Amp{f}^{\ast}\Pket{P} +\AmpB{f}^{\ast}\Pbarket{P}\Big] \,,
\end{equation}
and it is the one \emph{filtered} by the decay. Experimentally, the Reference transition $P_1\to P_2(t)$ is directly connected to $P_2=\nofp{g}$, i.e.
\begin{equation}
\nof{f}(t_1) \to \nofp{g}(t_2)\,. \label{eq:transition:00}
\end{equation}
Starting from \refeq{eq:transition:00}, the $T$ transformed transition
\begin{equation}
\nofp{g}(t_2) \to \nof{f}(t_1)\,. \label{eq:transition:01}
\end{equation}
does not correspond to the pair of decays $g$ at $t_1$ and $f$ at $t_2>t_1$, neither in the initial nor in the final decays. This is the ``orthogonality problem'', that prevents taking an arbitrary pair of decay channels.

To connect \refeq{eq:transition:01} with experiment, we need to find a pair of decay channels $f$, $f^\prime$, such that, for each of them,
\begin{equation}
\text{given } f,\quad \exists\ f^\prime\ /\  \Knof{f^\prime}=\Knofp{f}\,.\label{eq:condition:00}
\end{equation}
This condition is satisfied by either $CP$ conjugate decay channels $(\Pmeson{P},\Pbarmeson{P})$ or $CP$ eigenstates of opposite sign with the same flavour content\footnote{These decays should have a unique weak amplitude, i.e. no direct $CP$ violation.} $(P_+,P_-)$ \cite{Bernabeu:2012ab}. Hence the exceptionality of the transitions between semileptonic and $CP$ eigenstate decays \cite{Banuls:1999aj,Banuls:2000ki}. Furthermore they have a time evolution in the transition which does not need $\Gamma_{12}$. The method has also been applied to \KKmix\ at a $\Phi$ factory \cite{Bernabeu:2012nu}. As a consequence, the condition \refeq{eq:condition:00} limits the pair of decay channels suitable for $T$-symmetry tests \emph{if we start}, as a Reference, \emph{from the ``experimental'' transition} \refeq{eq:transition:00}. This limitation is the price to be paid to obtain a genuine $T$-violation asymmetry: given $f$, which is the decay channel $f^\prime$ such that the $P$ state not decaying to $f^\prime$ coincides with the orthogonal state to the $P$ state not decaying to $f$?

\noindent The precise general solution to this question in terms of the parameters defined below, in eq. \refeq{eq:DecayParams01}, is $\Li{f'}\Li{f}^\ast=-\abs{{q}/{p}}^2$. The solution adopted in reference \cite{Bernabeu:2012ab} is a particular case that is in complete agreement with this condition.
For the time-ordered decay products $(\ell^{+},J/\psi K_{S})$ the reference transition for the $B$ mesons is $\Pbarmeson{B}\to B_{-}$ so the T reverse transition $B_{-}\to \Pbarmeson{B}$ corresponds to the decay product detection of $(J/\psi K_{L},\ell^{-})$. Therefore in the asymmetry $A_{12}(t)$ -- in eq.(\ref{eq:asymmetry00}) -- $P_{1}=\Pbarmeson{B}$ and $P_{2}=B_{-}$.

The scientific community was recently interested in extending the $T$-symmetry tests to additional pairs of decay channels, in a program similar to the one developed for $CP$ violation studies. How to do it? Having identified the orthogonality problem, we give in this paper a bypass consisting in having an alternative reference transition once we fix the two channels $f$, $g$:
\begin{equation}
\nof{f}(t_1)\to \nof{g}(t_2)\,,\label{eq:transition:02}
\end{equation}
replacing \refeq{eq:transition:00}. In this case, the $T$-transformed transition corresponds to
\begin{equation}
\nof{g}(t_1)\to \nof{f}(t_2)\label{eq:transition:03}
\end{equation}
replacing \refeq{eq:transition:01}.

Whereas now the two initial $P$ states of eqs. \refeq{eq:transition:02} and \refeq{eq:transition:03} are directly connected to experiment, the price to be paid now is that the two final states are not. One has to work out what is the connection of the novel genuine theoretical $T$-asymmetry observable to experimental measurements.

As stated, the novel asymmetry proposed for a $T$-symmetry test is then
\begin{equation}
A(f,g;t)\equiv \frac{N\left( f,g;t\right) }{D\left(
f,g;t\right) }=\frac{\transPROB{\nof{f}}{\noft{g}}-\transPROB{\nof{g}}{\noft{f}}}{\transPROB{\nof{f}}{\noft{g}}+\transPROB{\nof{g}}{\noft{f}}}\,,\label{eq:Asymmetry:00}
\end{equation}
where $\transPROB{\nof{f}}{\noft{g}}$ is the probability that an initially prepared $\Knof{f}$ becomes, after a time $t$, a $\Knof{g}$ state. If $\Knoft{f}$ is the evolved state at time $t$ of a $\Knof{f}$ at time $t=0$, then this probability will be:
\begin{equation}
\transPROB{\nof{f}}{\noft{g}}=\abs{\BRAKET{\nof{g}}{\noft{f}}}^2\,.
\label{eq:TransProb01}
\end{equation}
The probability for the transition \refeq{eq:transition:02} can be rewritten, using closure in the two-dimensional space of the meson system
\begin{equation}
\transPROB{\nof{f}}{\noft{g}} = \BRAKET{\noft{f}}{\noft{f}} - \abs{\BRAKET{\nofp{g}}{\noft{f}}}^2\label{eq:TransProb:00}
\end{equation}
where the second term is directly connected to the probability for the decay of a tagged $\Knof{f}$ state at $t_1$ to $g$ after a time $t=t_2-t_1$, and given by the ``filtering identity''
\begin{equation}
\abs{\BRAKET{\nofp{g}}{\noft{f}}}^2=\frac{\abs{\BRA{g} W \Knoft{f}}^2}{\abs{\Amp{g}}^2+\abs{\AmpB{g}}^2}\,,\label{eq:FilterIdentity}
\end{equation}
where to first order in the weak decay hamiltonian $H_{w}$ and to all orders in strong interactions $W=U(\infty,0)H_w$. $U(\infty,0)$ is the strong evolution operator and is equal to the identity if we can neglect final state interactions. In this notation, $\Amp{f} =\BRA{f} W\Pket{P}$ and $\AmpB{f}=\BRA{f} W\Pbarket{P}$.

Note that the first term in the right hand side of \eq{eq:TransProb:00} is a well defined and measurable quantity: the ratio between the number of mesons that have not decayed after a time $t$ and the initial number tagged at $t_1$ as $\Knof{f}$ by the observation of the first decay $f$. We may then call this term the ``total survival probability''.

Similarly, for the transition \refeq{eq:transition:03} associated to the second term in the Motion Reversal Asymmetry \refeq{eq:Asymmetry:00} we have the results \refeq{eq:TransProb01}-\refeq{eq:FilterIdentity} with $g$ and $f$ reversed. 
For the same decay products $(\ell^{+},J/\psi K_{S}) $ as reference, the $B$ mesons states appearing in the first term of the asymmetry eq. (\ref{eq:Asymmetry:00}) are $\Pbarmeson{B}\to B_{+}$ so the T reverse term $B_{+}\to \Pbarmeson{B}$ corresponds to the observation of decay products $(J/\psi K_{S},\ell^{+}) $ plus the two total survival probabilities.
We emphasize that this observable becomes entirely measurable in a $1^{--}$ meson factory for any pair of decay channels $f,g$: one needs the probabilities (\ref{eq:FilterIdentity}) for the final decays and the two total survival probabilities. The last ingredient is avoidable in the numerator of the asymmetry by an appropriate cyclic combination of three decay channels $f_{1},f_{2},f_{3}$,
\begin{equation}
N_{3}(f_{1},f_{2},f_{3};t)=N(f_{1},f_{2};t)+N(f_{2},f_{3};t)+N(f_{3},f_{1};t).
\end{equation}
We have not imposed any particular condition to the pair of decay channels $f$, $g$, so one is not forced to use flavour specific or $CP$ eigenstate decay channels. We will discuss later whether the measurable \refeq{eq:Asymmetry:00} becomes a genuine Time Reversal Violation Asymmetry for any pair of decays.

Using the time evolution imposed by Quantum Mechanics we know the time dependent structure of the needed measurable quantities in terms of $\Delta m$  and $\Delta\Gamma$, determined by the eigenvalues of the entire Hamiltonian for the $\{\Pmeson{P},\Pbarmeson{P}\}$ system
\begin{equation}
\BRAKET{\noft{f}}{\noft{f}}=e^{-\Gamma t}\left\{\begin{array}{l}\srvCh{f}\tCh+\srvCc{f}\tCc\\ +\srvSh{f}\tSh+\srvSc{f}\tSc\end{array}\right\}\,,
\label{eq:TimeProb01}
\end{equation}%
\begin{equation}
\frac{\abs{\BRA{g}W\Knoft{f}}^2}{\abs{\Amp{g}}^2+\abs{\AmpB{g}}^2}=e^{-\Gamma t}\left\{\begin{array}{l}\trnCh{f}{g}\tCh+\trnCc{f}{g}\tCc\\ +\trnSh{f}{g}\tSh+\trnSc{f}{g}\tSc\end{array}\right\}\,.
\label{eq:TimeProb02}
\end{equation}
In other words, each one of the two transitions \refeq{eq:transition:02} and \refeq{eq:transition:03} is determined by the measurable parameters $\srvCh{f}$, $\srvCc{f}$, $\srvSh{f}$, $\srvSc{f}$ and $\trnCh{f}{g}$, $\trnCc{f}{g}$, $\trnSh{f}{g}$, $\trnSc{f}{g}$, not all of them independent. The numerator of the asymmetry \refeq{eq:Asymmetry:00} $N(f,g;t)$,
 is then given by three ``asymmetry parameters'': the non-vanishing value of any of these asymmetry parameters would be a signal of Time Reversal Violation.
\begin{equation}
N(f,g;t)=e^{-\Gamma\, t}\left\{\begin{matrix}\numC{f}{g}\left[\tCh-\tCc\right]\\ +\numSh{f}{g}\tSh+\numSc{f}{g}\tSc\end{matrix}\right\}\,.\label{eq:TimeProbCoef03}
\end{equation}
Notice that the asymmetry parameters are obtained as a difference between the total survival probability \refeq{eq:TimeProb01} and the filtering identity \refeq{eq:TimeProb02}. 

The theoretical connection of these measurable parameters with the familiar quantities involved in the Weisskopf-Wigner approach (WWA) \cite{Weisskopf:1930au,Weisskopf:1930ps,Lee:1957qq} for the $\{\Pmeson{P},\Pbarmeson{P}\}$, in terms of the Hamiltonian matrix elements $H_{ij}$, is
\begin{eqnarray}
\numC{f}{g}&=& \delta(\Cf-\Cg)-\im{\theta}(\Sf-\Sg)\,,\label{eq:TimeProbCoef03-1}\\
\numSh{f}{g}&=& \delta(\Cf\Rg-\Cg\Rf)+\im{\theta}(\Rf\Sg-\Rg\Sf)\,,\label{eq:TimeProbCoef03-2}\\
\numSc{f}{g}&=& (\Cf\Sg-\Cg\Sf)(1+\delta(\Cf+\Cg))\notag\\
	    && +\delta(\Sf-\Sg)+\re{\theta}(\Rf\Sg-\Rg\Sf)\,.\label{eq:TimeProbCoef03-3}
\end{eqnarray}
where we have presented the results at leading order in terms of the usual mixing parameters \cite{Branco:1999fs} $\delta =\left(1-\abs{q/p}^{2}\right)/\left(1+\abs{q/p}^{2}\right)$ a real number responsible of $CP$ and $T$  violation in $H_{12}$, and the complex parameter $\theta$ that is $CPT$ and $CP$ violating, $\theta =\left(H_{22}-H_{11}\right)/\left(\Delta m-i\Delta \Gamma /2\right)$. Including the weak decay transition, we also use the parameters $\Ri{i}$, $\Si{i}$ and $\Ci{i}$ defined in terms of the well known ``mixing $\times$ decay'' interference quantities $\Li{i}=q\AmpB{i}/p\Amp{i}$,
\begin{equation}
\Ci{i}=\frac{1-\abs{\Li{i}}^2}{1+\abs{\Li{i}}^2}\,,\quad \Si{i}=\frac{2\,\im{\Li{i}}}{1+\abs{\Li{i}}^2}\,,\quad \Ri{i}=\frac{2\,\re{\Li{i}}}{1+\abs{\Li{i}}^2}\,,\label{eq:DecayParams01}
\end{equation}
where $q/p$ is defined as $(q/p)^2=H_{12}/H_{21}$, even for the $CPT$ violating case $\theta\neq 0$. Note that these parameters are not all independent: $\Ci{i}^{2}+\Si{i}^{2}+\Ri{i}^{2}=1$.

\noindent In order to show how the new $T$-violating observable opens the $T$ analysis to more channels at a B-factory, we will concentrate in the case where one decay channel $f$ is a flavour specific channel corresponding to $\Knof{f}=\Pbarket{B}$ ($\Knof{f}=\Pket{B}$) and $\AmpB{f}=0$ ($\Amp{f}=0$), therefore
\begin{equation}
\Ci{f}=1\, (\Ci{f}=-1)\,;\quad \Si{f}=0\,;\quad \Ri{f}=0\,.
\end{equation}
For a B-factory it is an excellent approximation to put $\Delta\Gamma=0$ and $\delta=0$. Under these conditions, we have for any other channel $g$ the numerator of the asymmetry
\begin{align}
&N(\Pmeson{B},g;t)=-N(g,\Pmeson{B};t) = e^{-\Gamma t}\Si{g}\left\{\tSc+\im{\theta}[1-\tCc]\right\}\,,\hfill\notag\\
&N(\Pbarmeson{B},g;t)=-N(g,\Pbarmeson{B};t) = e^{-\Gamma t}\Si{g}\left\{-\tSc+\im{\theta}[1-\tCc]\right\}\,.\label{eq:NumeratorsB:00}
\end{align}
The corresponding denominators for these transitions are
\begin{align}
&D(\Pmeson{B},g;t)=D(g,\Pmeson{B};t)=\hfill\notag\\
\hfill &e^{-\Gamma t}\left\{1+\Ci{g}\tCc-\Ri{g}\re{\theta}[1-\tCc]+\im{\theta}(1+\Ci{g})\tSc\right\}\,,\notag\\
&D(\Pbarmeson{B},g;t)=D(g,\Pbarmeson{B};t)=\hfill\notag\\
\hfill &e^{-\Gamma t}\left\{1-\Ci{g}\tCc+\Ri{g}\re{\theta}[1-\tCc]-\im{\theta}(1+\Ci{g})\tSc\right\}\,,\label{eq:DenominatorsB:00}
\end{align}
A superficial inspection of \refeq{eq:NumeratorsB:00} would indicate that the existence of a non-vanishing factor $\Si{g}$ is, by itself, a signal of $T$-violation. This conclusion would be wrong, in general. The problem in quantum mechanics can be traced back to the antiunitary character of the operator $U_T$ implementing the $T$ transformation in the space of physical states. The principle of $T$-invariance or microreversibility strictly means 
\begin{equation}
\abs{\BRA{U_T\nof{f}}U_2(t,0)\KET{U_T\nof{g}}}^2=\abs{\BRA{\nof{g}}U_2(t,0)\KET{\nof{f}}}^2\,,
\end{equation}
with $U_T=\widehat U_T K$, where $\widehat U_T$ is unitary and $K$ is the complex conjugation operator in a definite basis. For pseudoscalars, the effect of $\widehat U_T$ is safe with the only change $\vec p\to -\vec p$ and using rotational invariance. The possible difficulty in asigning a genuine character of $T$-violation to the asymmetry \refeq{eq:Asymmetry:00} is concentrated on $K$. Sufficient conditions to satisfy
\begin{equation}
U_T\Knof{g}=e^{i\varphi}\Knof{g}\label{eq:Condition:00}
\end{equation}
in the $T$-invariant limit will be discussed in a separate publication \cite{nuevo}. Here we mention an incomplete list of types of decay channels that can be used with the effect of $U_T$ being a global phase factor as in \eq{eq:Condition:00}:
\begin{enumerate}
\item Flavour Specific channels.
\item $CP$ eigenstates without direct $CP$ violation. The absence of direct $CP$ violation in the decay to $CP$ eigenstates makes irrelevant the presence of the complex conjugation operator, in the sense that it changes the global phase of $\Knof{i}$, and therefore $\Si{g}$ is a genuine signal of $T$-violation.
\item For non $CP$ eigenstates without $CP$ violation in the decay, one can prove \cite{nuevo}, with $\xi$ given by $U_{CP}\Pket{P}=e^{i\xi}\Pbarket{P}$ and $U_{CP}$ the unitary operator implementing the $CP$ transformation,
\begin{equation}
\Li{g}\Li{\bar g}=e^{-i2\xi}\left(\frac{q}{p}\right)^2\,,
\end{equation}
in such a way that $\Li{g}\Li{\bar g}\neq 1$ is a signal of $T$ violation. As a consequence, the genuine signal of $T$ violation is given by the condition $\Si{g}+\Si{\bar g}\neq 0$.
\item A decay product which is an eigenstate of the Strong Scattering Matrix. One case of possible interest at a $\Phi$ factory is that of the $\Pmeson{K},\Pbarmeson{K}\to (\pi\pi)_I$ channels with definite isospin $I$.
\end{enumerate}
It is worth noting that the asymmetries constructed with the combination of semileptonic and $CP$ eigenstate decay channels are now different from those already measured by the BABAR Collaboration \cite{Lees:2012uka,Bernabeu:2012ab,Banuls:1999aj,Banuls:2000ki}. A case of particular interest is given by the Reference transition $\Pmeson{B}$, or $\Pbarmeson{B}$, decaying to $J/\Psi K_S$, which does not involve in the novel asymmetry the decay channel $J/\Psi K_L$. This case can be extended to other charmonium-$K_S$ final states, and also to $\phi K_S$, $\pi^0 K_S$, $\rho^0K_S$, $J/\Psi \pi^0$ and $(\pi\pi)_{I=2}$ \cite{Baek:2005cg} as interesting examples.

For decay products which are not $CP$ eigenstates, our result suggests the use of $N(\Pmeson{B},g;t)$ and $N(\Pmeson{B},\bar g;t)$, which can be combined to avoid any possible presence of fake $T$ violation in the individual transitions. The following observable
\begin{multline}
N(\Pmeson{B},g;t)+N(\Pmeson{B},\bar g;t)=\\
e^{-\Gamma t}\,(\Si{g}+\Si{\bar g})\left\{\tSc+\im{\theta}(1-\tCc)\right\}
\end{multline}
for decay channels $g$, $\bar g$ without direct $CP$ violation is a genuine $T$ violation observable. Among those channels we have, for example, $g=D^\ast(2010)^+D^-$, $\bar g=D^\ast(2010)^-D^+$, where this asymmetry should be almost maximal: $\Si{g}+\Si{\bar g}=2\times (-0.73\pm 0.11)$ \cite{Amhis:2012bh}. Other channels like $D^{(\ast)\pm}\pi^\mp$ or even $\Pmeson{B_s}\to D_s^\pm K^\mp$ should be interesting.
%

To summarize, we have demonstrated the possibility of extending the $T$ violation tests to more decay channels by means of a novel strategy: instead of searching which is the pair of decay channels associated to the $T$-reverse meson transition, we build an asymmetry which automatically tags the initial states of both the Reference and the $T$-reverse meson transitions. The connection to the experimental quantities to be measured requires the determination of the total survival probability of the meson, as well as the filtering probability for the considered decay. The proposed asymmetry can be used for decay products which are not $CP$ eigenstates by combining the observables for the Reference transitions $\Pmeson{B}\to g$ with $\Pmeson{B}\to\bar g$. In this way we eliminate possible sources of fake $T$ violation. 
With the proposal discussed in this letter, the way is open to a full experimental programme of studies of $T$ violation observables at meson factories \cite{Bevan:2013rpr}, using the quantum principles of entanglement and the decay as a filtering measurement.
%
%
\section*{Acknowledgments\label{SEC:Ack}}
The authors thank Pablo Villanueva and Fernando Mart\'\i nez-Vidal for interesting conversations on the subject of this paper. F.J.B. acknowledges illuminating discussions with Jo\~ao Silva. 
\noindent This work was supported by Spanish MINECO under grant FPA2011-23596 and by \emph{Generalitat Valenciana} under grants PROMETEO 2010-056 and PROMETEOII 2013-017.
%
\section*{Bibliography\label{SEC:bib}}

\end{document}